\documentclass[aps,prb,twocolumn,superscriptaddress]{revtex4-1}

\usepackage{graphicx}
\usepackage{dcolumn}
\usepackage{bm}
\usepackage{subfigure}
\usepackage{wrapfig}
\usepackage{amssymb,latexsym, amsmath}
\usepackage{graphics}

\begin{document}


\title{Field-Tuned Superconductor-Insulator Transition in BaPb$_{1-x}$Bi$_x$O$_3$}


\author{P. \surname{Giraldo-Gallo}}
\affiliation{Geballe Laboratory for Advanced Materials, Stanford University, Stanford, CA 94305, USA}
\affiliation{Department of Physics, Stanford University, CA 94305, USA}
\author{Hanoh Lee}
\affiliation{Geballe Laboratory for Advanced Materials, Stanford University, Stanford, CA 94305, USA}
\affiliation{Department of Applied Physics, Stanford University, CA 94305, USA}
\author{Y. Zhang}
\author{M. J. Kramer}
\affiliation{Ames Laboratory (USDOE), Department of Materials Science and Engineering, Iowa State University, Ames IA 50011-3020, USA}
\author{M. R. Beasley}
\author{T. H. Geballe}
\author{I. R. Fisher}
\affiliation{Geballe Laboratory for Advanced Materials, Stanford University, Stanford, CA 94305, USA}
\affiliation{Department of Applied Physics, Stanford University, CA 94305, USA}


\date{\today}

\begin{abstract}
BaPb$_{1-x}$Bi$_x$O$_3$ is found to exhibit a field-tuned superconductor to insulator transition for Bi compositions 0.24 $\leq x \leq$ 0.29. The magnetoresistance of optimally doped samples manifests a temperature-independent crossing point and scaling of the form $\rho(T,H)=\rho_c F( |H-H_{c}|T^{-1/z\nu})$, where $H_c$ is the field determined by the temperature-independent crossing point, and $z\nu$ = 0.69 $\pm$ 0.03. High resolution transmission electron microscopy measurements reveal a complex intergrown nanostructure comprising tetragonal and orthorhombic polymorphs. Data are analyzed in terms of both a classical effective medium theory and a field-tuned quantum phase transition, neither of which provides a completely satisfactory explanation for this remarkable phenomenology.
\end{abstract}


\maketitle

\section{Introduction}

BaBiO$_3$ is a charge density wave (CDW) insulator comprising two distinct Bi sites.\cite{Varma1, Tajima2, Cox1, Franchini1, Yin1} Chemical substitution of either K for Ba or Pb for Bi results in suppression of the CDW state, and for a limited range of compositions yields a superconductor, with maximum $T_c$ values of 12 and 30 K for BaPb$_{1-x}$Bi$_x$O$_3$ and Ba$_{1-x}$K$_x$BiO$_3$ respectively.\cite{Sleight1, Cava2, Gabovich1, Uchida1, Suzuki1, Tarapheder1} Given their moderately high $T_c$ values, these materials have attracted considerable attention in order to determine the superconducting pairing interaction, and factors determining the phase diagram. Significantly, chemical substitution not only modifies the electronic structure,\cite{Yin1} but also introduces disorder, and with it a whole new set of issues that can play an important role in both the superconducting properties and also the compositionally-tuned metal-insulator transition. 

Several previous studies of BaPb$_{1-x}$Bi$_x$O$_3$ including x-ray diffraction,\cite{Khan1, Sleight2, Cava1} transport measurements,\cite{Suzuki1} optical reflectivity,\cite{Tajima1, Puchkov1} and thermodynamic measurements,\cite{Wolf1} have provided important insights to the evolution of the crystal structure and physical properties across the phase diagram. The material superconducts for compositions $0.05\lesssim x \lesssim 0.35$, but the volume fraction is always less than 100 \%. Recent high resolution x-ray and neutron scattering measurements have revealed the presence of two polymorphs in this compositional range, one with a tetragonal symmetry and one orthorhombic.\cite{Cava1} The superconducting volume fraction scales with the relative proportion of the tetragonal phase, leading to the conclusion that only the tetragonal phase harbors superconductivity. However, it is not clear what factors determine $T_c$ in this material, and previous studies have neglected the profound effects of localization and enhanced interactions arising from disorder. 

In this paper we present magnetoresistance measurements of single crystal samples of superconducting BaPb$_{1-x}$Bi$_x$O$_3$ that reveal a remarkable temperature-independent crossing point for compositions close to optimal doping. To account for this phenomenology within an effective medium theory (used to account for the dimorphic nature of the material in this compositional range) requires a possibly unphysical degree of fine-tuning. On the other hand, the existence of a temperature-independent crossing point also motivates analysis of the data within the context of a field-tuned quantum phase transition (QPT). The data scales remarkably well for an effective dimensionality $d=2$, yielding a product of critical exponents of $z\nu=0.69\pm 0.03$. It is not clear why the resistivity should scale in this manner for this fundamentally 3-dimensional material, though the result is probably related to the structural dimorphism revealed by our high resolution TEM measurements. This surprising phenomenology exposes the need for a theory incorporating aspects of both percolation and quantum critical fluctuations. Our results also imply a key role for disorder in determining $T_c$ in this material, perhaps also accounting for the lower maximum $T_c$ of BaPb$_{1-x}$Bi$_x$O$_3$ relative to that of Ba$_{1-x}$K$_x$BiO$_3$. 

\section{Experimental Methods}

Single crystals of BaPb$_{1-x}$Bi$_x$O$_3$ with 0 $\leq x \leq$ 0.29 were grown using a self-flux technique, similar to that described in Ref.  \onlinecite{Katsui1}. The cation composition was determined by electron microprobe analysis. These measurements revealed a uniform composition across each sample, within the experimental uncertainty ($\pm$ 0.02). $T_c$ and the superconducting volume fraction were characterized by measurements of the field-cooled and zero-field-cooled susceptibility using a commercial Quantum Design SQuID magnetometer.

Resistivity measurements were made using a standard four-probe configuration. Crystals were initially cut and cleaved to obtain approximately rectilinear bars with an arbitrary orientation with respect to the crystal axes. Electrical contact was made using sputtered gold pads. To minimize uncertainty in absolute values of the resistivity due to geometric factors, at least four samples for each composition were measured and the average value used to scale the curves that are shown. 

High resolution transmission electron microscopy (HRTEM) measurements were taken for samples with bismuth composition of $24\%$. Samples were crushed in liquid nitrogen cooled ethyl alcohol, and the liquid was allowed to warm to room temperature.  The slurry was stirred and a small droplet was placed on a holey carbon grid and dried in air.  The sample was analyzed using a FEI G2 F20 Tecnai STEM operated at 200 keV.  Thin areas were analyzed with selected area diffraction, energy dispersive spectroscopy and high resolution imaging.  Thin areas were aligned with the $\left[010\right]$ zone axis based on indexing to the Ibmm structure (space group $\#74$), showing clear lattice fringes in the HRTEM.

\begin{figure}
\vspace{-0.7cm}
\includegraphics{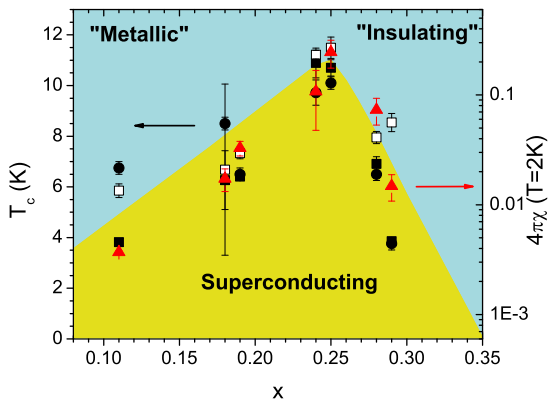}\vspace{-0.5cm}
\caption{(Color online) Variation of the critical temperature $T_c$ (left axis) and zero-field-cooled magnetic susceptibility (right axis, red triangles) of single crystals of BaPb$_{1-x}$Bi$_x$O$_3$ as a function of the Bi concentration $x$. $T_c$ values were obtained from measurements of resistivity (open black squares for a 90$\%$ criteria, and filled black squares for a 50 $\%$ criteria) and magnetic susceptibility (black filled circles). }\label{fig_phaseDiag}
\vspace{-0.3cm}
\end{figure}

\begin{figure}
\vspace{-0.3cm}
\includegraphics{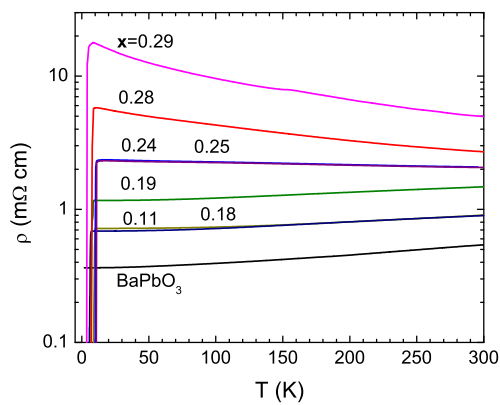}\vspace{-0.5cm}
\caption{(Color online) Temperature dependence of the resistivity of BaPb$_{1-x}$Bi$_x$O$_3$. Curves are labeled according to the Bi concentration, $x$.}\label{fig_rhoT}
\vspace{-0.25cm}
\end{figure}

\begin{figure*}[ht]
\begin{minipage}[b]{0.4\linewidth}
\vspace{-1cm}
\centering
\includegraphics[scale=1]{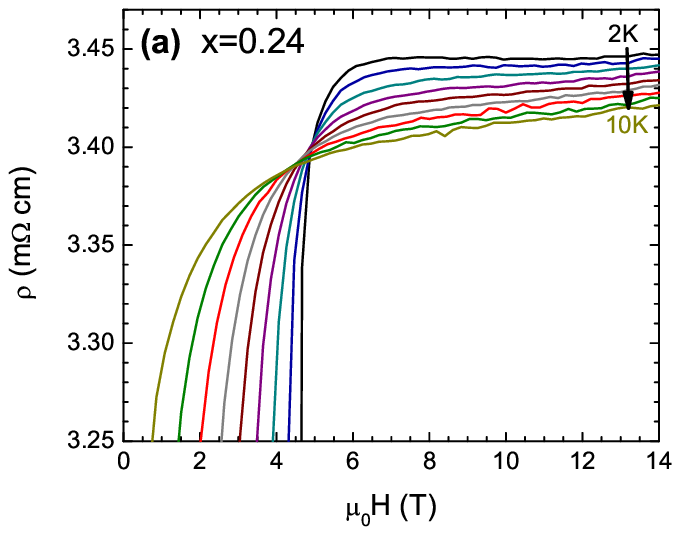}
\end{minipage}
\hspace{0.5cm}
\begin{minipage}[b]{0.4\linewidth}
\vspace{-1cm}
\centering
\includegraphics[scale=1]{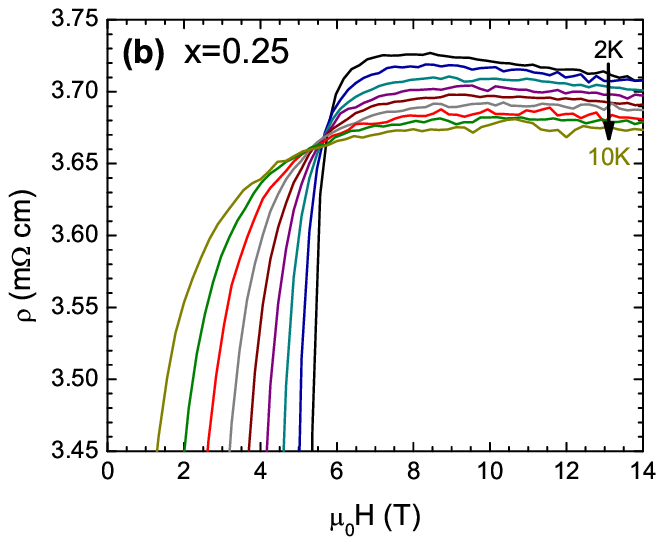}
\end{minipage}
\hspace{0.1cm}
\begin{minipage}[t]{0.4\linewidth}
\vspace{-1cm}
\centering
\includegraphics[scale=1]{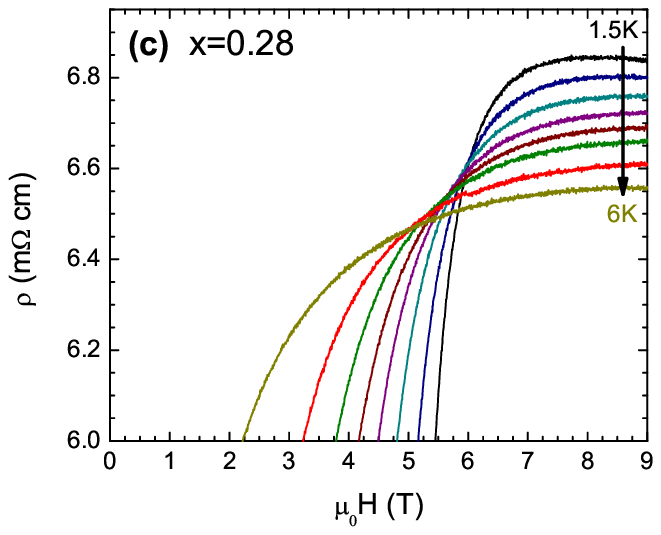}
\end{minipage}
\hspace{0.5cm}
\begin{minipage}[t]{0.4\linewidth}
\vspace{-1cm}
\centering
\includegraphics[scale=1]{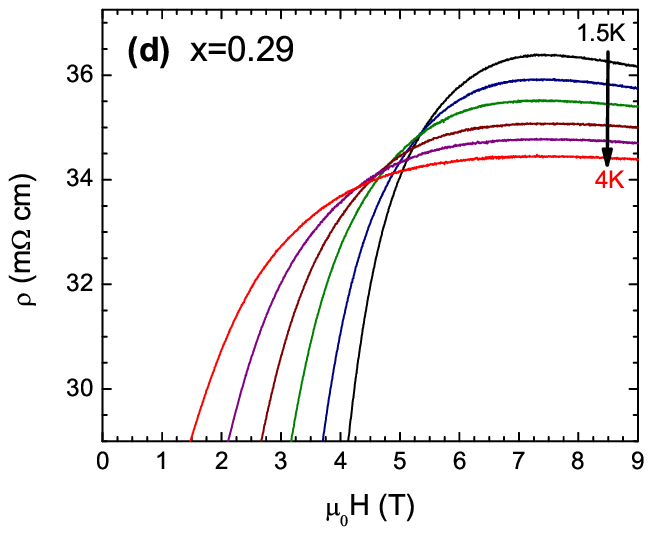}
\end{minipage}
\vspace{-0.4cm}
\caption{(Color online) Isotherms of resistivity vs. magnetic field for bismuth concentrations of {\bf (a)} $x$=0.24 (2 K to 10 K in 1 K steps), {\bf (b)} $x$=0.25 (2 K to 10 K in 1 K steps), {\bf (c)} $x$=0.28 (1.5 K, 2 K, 2.5 K, 3 K, 3.5 K, 4 K, 5 K and 6 K), {\bf (d)} $x$=0.29 (1.5 K to 4 K in 0.5 K steps). For the nearly optimally doped compositions $x$ = 0.24 and 0.25 a temperature-independent crossing point is evident for temperatures below approximately 8 K. }\label{fig_rhoH}
\vspace{-0.2cm}
\end{figure*}

\begin{figure}
\vspace{-0.4cm}
\includegraphics[scale=0.95]{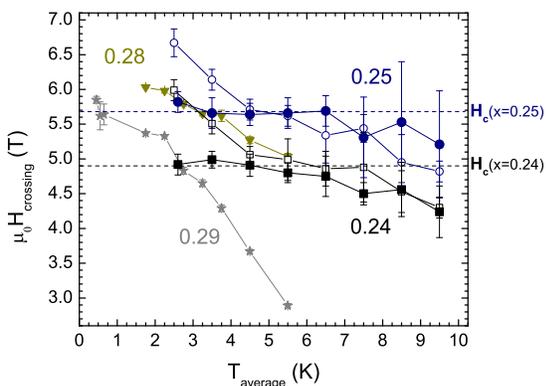}
\vspace{-0.5cm}
\caption{(Color online) Temperature dependence of the crossing field $H_{crossing}$ for adjacent isotherms, obtained from magnetoresistance measurements. $T_{ave}$ is the average temperature. Compositions that exhibit a temperature independent crossing point $H_c$ ($x=0.24$ and $0.25$) are shown by black solid squares and blue solid circles respectively. Open symbols show low temperature deviation from $H_c$ for some samples measured. For higher Bi concentrations ($x=0.28$ and $0.29$; triangles and stars respectively) the crossing field does not converge to a single value over any range of temperature.}\label{fig_Hcrossing}
\end{figure}

\begin{figure*}
\begin{center}
\vspace{-0.3cm}
\hspace{-0.1cm}
\centering
\subfigure{
\includegraphics[scale=0.265]{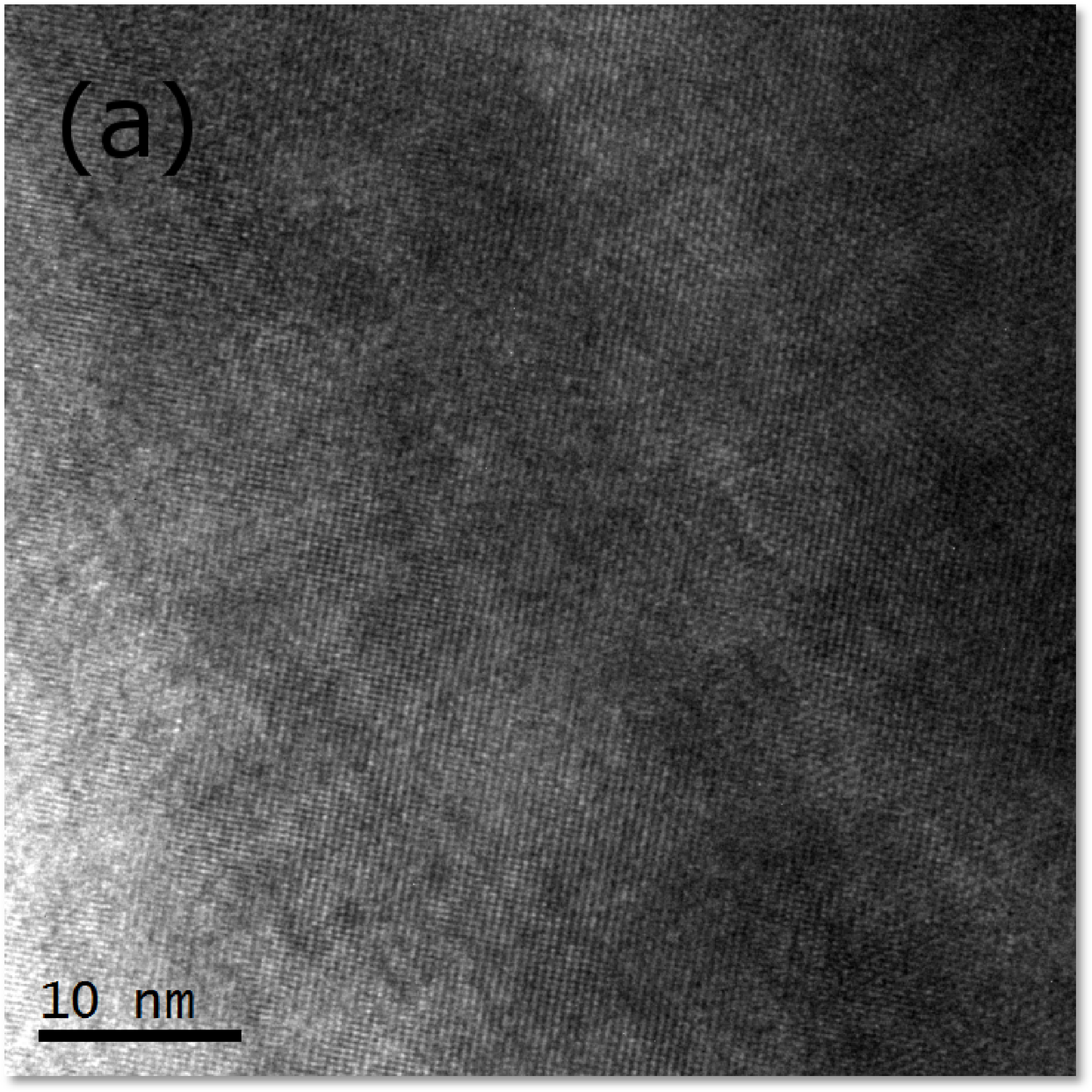}
}
\hspace{1cm}
\subfigure{
\includegraphics[scale=0.465]{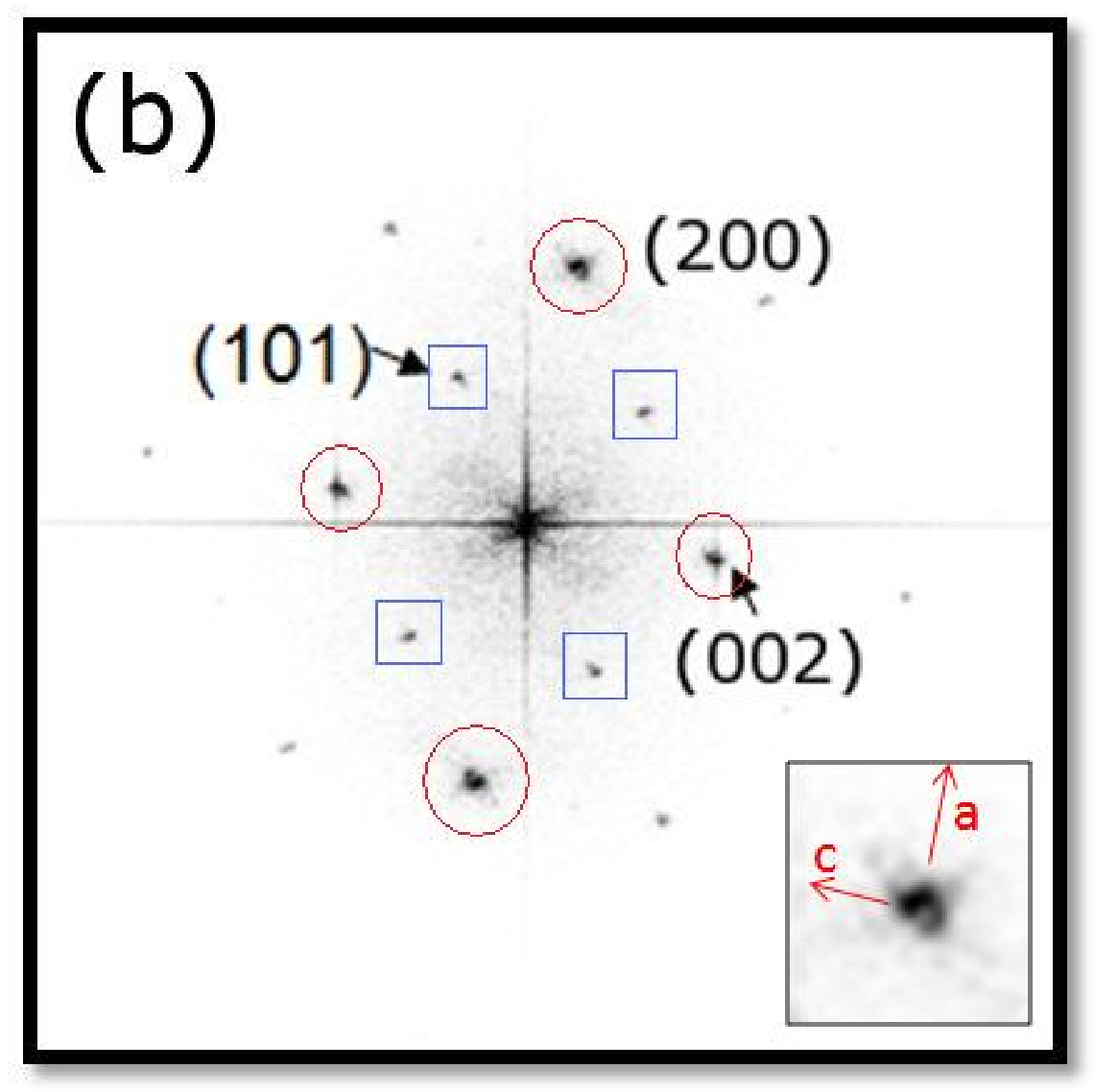}
}
\vspace{-0.5cm}
\end{center}
\caption{(Color online) {\bf (a)} HRTEM image for a sample with Bi composition of $x$=0.24, seen along the $\left[010\right]$ zone axis. {\bf (b)} Fast Fourier transform (FFT) of the HRTEM revealing the weaker $\{101\}_T$ reflections (in blue squares) attributed to the Ibmm orthorhombic phase as well as the $\{200\}_T$ reflections (in red circles) from both tetragonal and orthorhombic phases. Inset shows an expanded view to the $(200)$ peak, where streaks in the $\left[102\right]_T$ and $\left[-102\right]_T$ directions can be seen.}\label{fig_HRTEM}
\end{figure*}

\begin{figure*}
\begin{center}
\hspace{-0.1cm}
\centering
\subfigure{
\includegraphics[scale=0.25]{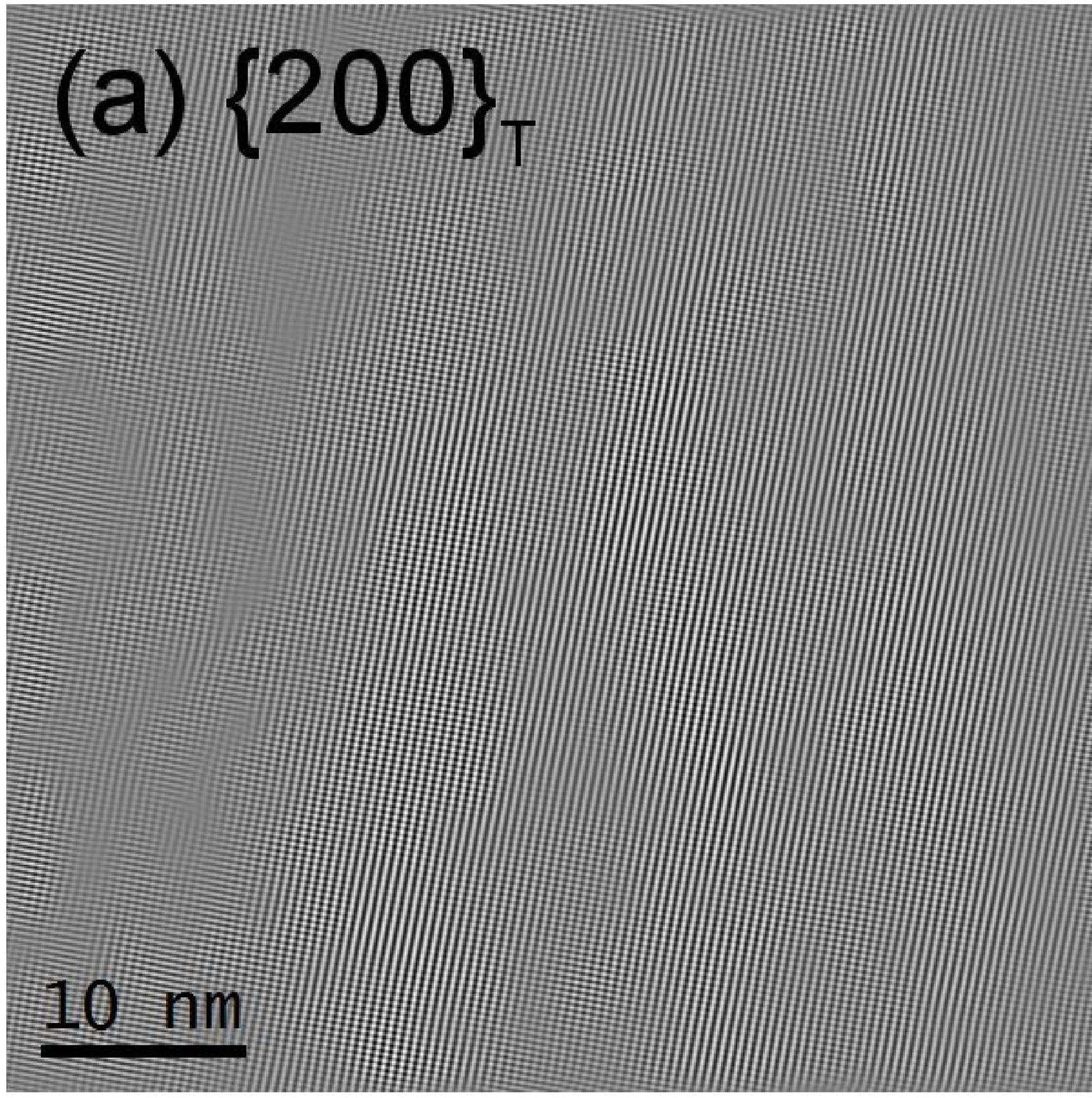}
}
\hspace{0.2cm}
\subfigure{
\includegraphics[scale=0.25]{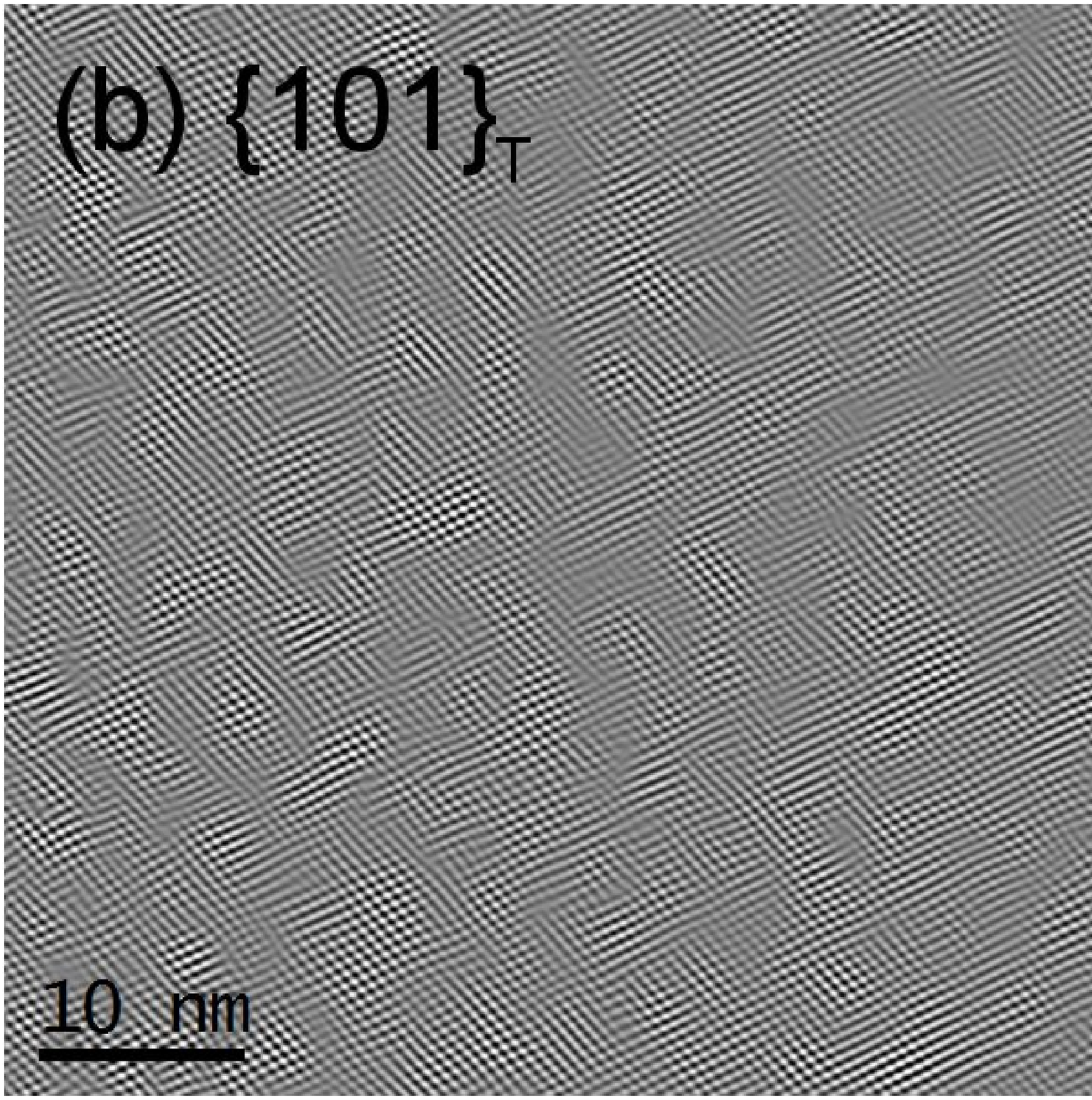}
}
\hspace{0.2cm}
\subfigure{
\includegraphics[scale=0.25]{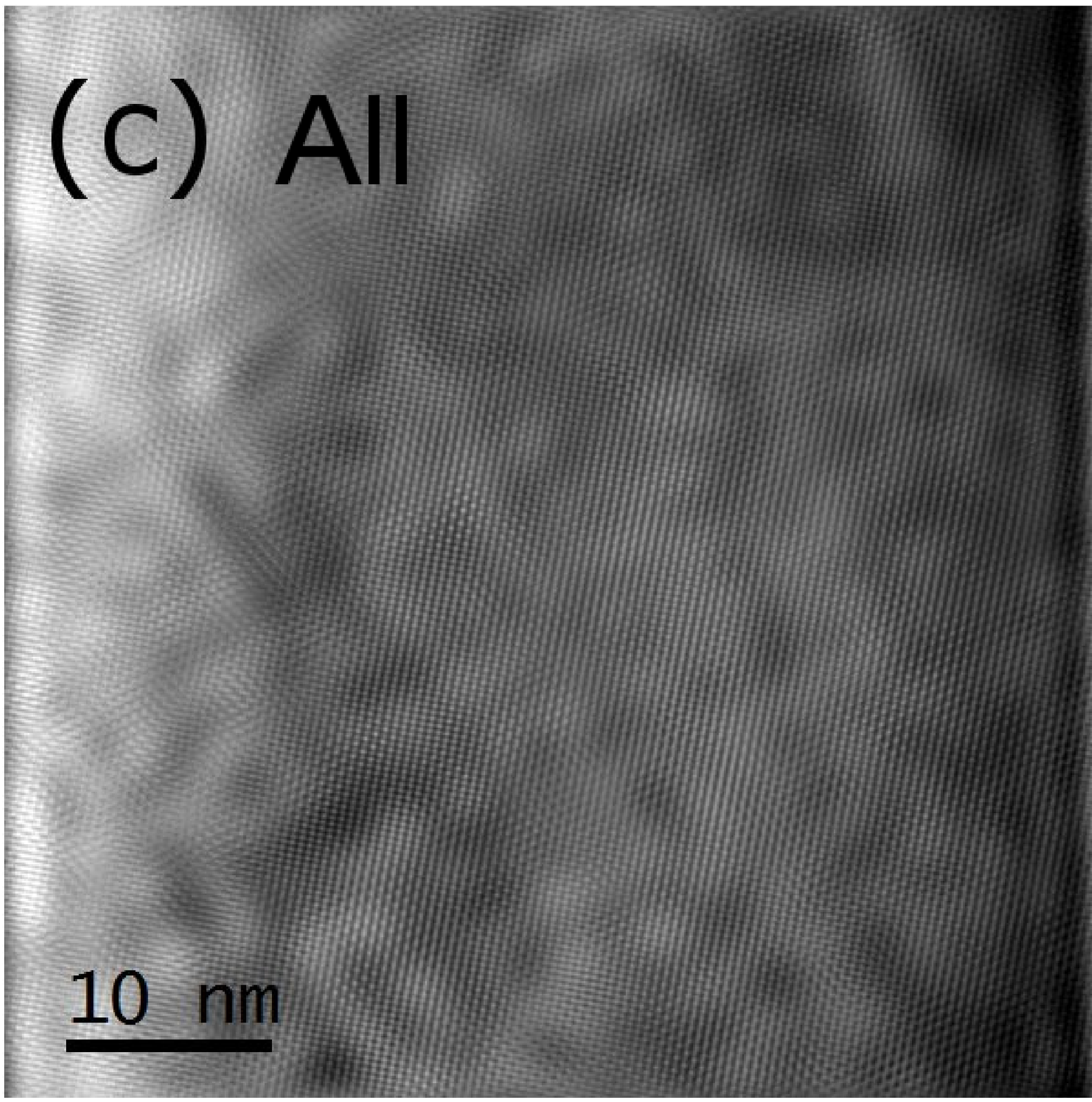}
}
\vspace{-0.5cm}
\end{center}
\caption{Images reconstructed from the HRTEM after Fourier filtering via masks applied to the {\bf (a)} $(200)_T$ and $(002)_T$ main reflections, and {\bf (b)} $(101)_T$ weaker reflections of the FFT shown in Fig. \ref{fig_HRTEM}(b). {\bf (c)} Filtered enhanced image resulting from including both kind of reflections.}\label{fig_TEM}
\end{figure*}

\section{Results}

$T_c$ values obtained by magnetic susceptibility measurements were estimated using a $1\%$ criteria (0.01 of the value at the base temperature of 2 K). The low-temperature value of the zero-field-cooled susceptibility (4$\pi\chi(T$=2K)) was used to provide an indication of the superconducting volume fraction. A maximum value of $\thicksim 25\%$ was obtained for the optimally doped samples ($x\approx$0.25), this value decreasing rapidly for lower or higher Bi concentrations (see Fig. \ref{fig_phaseDiag}). This behavior is consistent with results previously found in polycrystalline samples \cite{Cava1}, though with a reduced maximum volume fraction, possibly due to the different conditions and cooling rates used to grow single crystal samples.

Figure \ref{fig_rhoT} shows curves of electrical resistivity vs. temperature at zero field for representative samples of different Bi compositions. $T_c$ values obtained from these measurements (using both 50 and 90$\%$ criteria), are shown in Fig. \ref{fig_phaseDiag} and agree with values obtained from susceptibility measurements. Low Bi concentrations show ``metallic-like'' resistivity, with $\partial\rho/\partial T > 0$, but very high absolute values of the resistivity, whereas concentrations of 24$\%$ and above show ``insulating-like'' behavior, with $\partial\rho/\partial T< 0$. The separatrix between insulating and metallic behavior appears to occur very close to optimal doping and before the composition where the gap associated with the CDW opens ($x\approx 0.35$).\cite{Tajima1}

Magnetoresistance measurements were made for fields up to 14 T oriented perpendicular to the current. Curves of resistivity as a function of magnetic field for different temperatures were taken for multiple samples from each of the compositions studied. The data were symmetrized from positive and negative field sweeps, although the Hall component was very small for all insulating compositions. For metallic-like samples ($x \leq$ 0.19), the resistive transition is uniformly suppressed for increasing field or temperature, as anticipated. Results for the insulating-like samples ($x \geq$ 0.24) are qualitatively different. Representative data for four different compositions with progressively larger Bi concentrations are shown in Fig. \ref{fig_rhoH}. In the presence of a finite magnetic field, these samples show a clear crossover from a superconducting-like resistivity (decreasing resistivity with decreasing temperature) to an insulating-like one (increasing resistivity with decreasing temperature). Remarkably, the samples closest to optimal doping ($x \approx$ 0.24 - 0.25) show a temperature-independent crossing point. Plotting the crossing field $H_{crossing}$ for adjacent isotherms as a function of the average temperature ($T_{ave}$) reveals the convergence to a ``critical'' magnetic field $H_{c}$ (Fig. \ref{fig_Hcrossing}) within the resolution set by the signal-to-noise ratio of individual field sweeps. The range of temperatures over which this crossing point is observed to be temperature independent varies slightly from sample to sample. For the examples shown in Figs. \ref{fig_rhoH}(a) and \ref{fig_rhoH}(b) it is found for temperatures from approximately 8 K down to 2 K, the lowest temperature to which these samples were measured (solid symbols in Fig. \ref{fig_Hcrossing}). Other samples with the same nominal composition reveal a similar crossing point from 8 K down to just 3 or 4 K (open symbols in Fig. \ref{fig_Hcrossing}), below which $H_{crossing}$ deviates from $H_c$. In contrast, for samples with a somewhat higher bismuth concentration of $x =$ 0.28 and 0.29 (Fig. \ref{fig_rhoH}(c) and (d)), the crossover from the superconducting to the insulating state is not demarcated by a temperature-independent crossing point over any range of temperatures, and $H_{crossing}$ in Fig. \ref{fig_Hcrossing} does not converge to a single value.

To gain insight to the origin of the unusual magnetoresistance, HRTEM measurements were made for samples with a bismuth concentration of $x$=0.24. Representative results are shown in Fig. \ref{fig_HRTEM}. The high resolution image (Fig. \ref{fig_HRTEM}(a)) reveals a well-ordered structure. A fast Fourier transform (FFT) of this image (Fig. \ref{fig_HRTEM}(b)) reveals peaks from both tetragonal ($hkl$ even) and orthorhombic ($hkl$ even and odd, in the tetragonal notation) phases. By systematically masking these diffraction peaks and performing an inverse Fourier transform it is possible to recreate the spatial separation of these two distinct polymorphs (Fig. \ref{fig_TEM}). The result of applying a mask such that only the $(200)_T$ and $(002)_T$ peaks common to both the tetragonal and orthorhombic phases are considered is shown in Fig. \ref{fig_TEM}(a).  An ordered array of planes of atoms is evident, indicating that the two structural phases are highly coherent through the crystal lattice. In contrast, the result of applying a mask to the $(101)_T$ peaks attributed only to the Ibmm orthorhombic phase, shown in Fig. \ref{fig_TEM}(b), reveals a spatial variation due to the densely intergrown nanostructure. The clusters are approximately 5-10 nm in size, which is comparable to the coherence length of 8 nm calculated from $H_{c2}(0)$. Finally, Fig. \ref{fig_TEM}(c) shows the result of applying a mask that considers both groups of reflections ($\{n00\}_T$ and $\{00n\}_T$ with $n=2,4,...$; and $\{m0m\}_T$ with $m=1,2,...$), more clearly revealing the nanostructure. The material appears to comprise a network of interpenetrating orthorhombic and tetragonal phases. 

In addition to the diffraction peaks, weak streaks are also evident in the FFT shown in Fig. \ref{fig_HRTEM}(b) running in the $\left[102\right]_T$ and $\left[-102\right]_T$ directions. The streaking is weaker for peaks associated purely with the orthorhombic phase, so is tentatively ascribed to a property primarily of the tetragonal clusters. These features are indicative of small domains whose interfaces are orthogonal to the beam (in the plane of the sample). While quantifying these is very difficult, the qualitative interpretation is straight forward; the diffuse pattern, i.e., the streak, is orthogonal to the interface. In the plane of the image, the tetragonal regions therefore correspond to elongated domains with the long axis oriented along the $\left[201\right]_T$ and $\left[-201\right]_T$ directions. These regions are evident in Fig. \ref{fig_TEM}(c) as extended regions of light contrast running approximately from top left to bottom right and bottom right to top left of the image. It is not clear, however, what the full 3D geometry of these regions is, or their connectivity in the b-axis direction.

\section{Discussion}\label{discuss}

The apparent temperature-independent crossing point observed for nearly optimally doped compositions is remarkable and requires explanation. The granular nanostructure revealed by HRTEM measurements suggests two initial avenues to understand the experimental results, based first on a classical effective medium theory, and second on a field-tuned quantum phase transition. We discuss each of these approaches below.

\begin{figure*}
\begin{center}
\vspace{-1cm}
\hspace{-0.9cm}
\centering
\subfigure{
\includegraphics[scale=0.95]{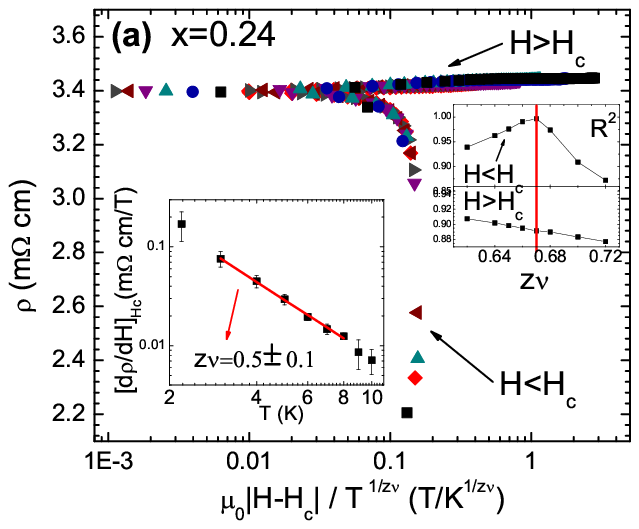}
}
\subfigure{
\includegraphics[scale=0.95]{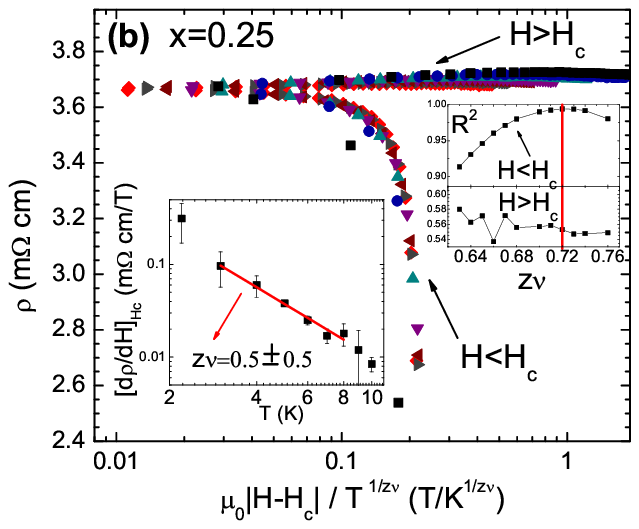}
}
\vspace{-0.8cm}
\end{center}
\caption{(Color online) Scaling analysis of representative samples with bismuth composition of {\bf (a)} $x$=0.24, {\bf (b)} $x$=0.25. $z\nu$ was initially calculated from a fit to $(d\rho/dH)|_{Hc}$ vs $T$ (lower left inset to each panel), and refined based on regression of the entire data set (upper right inset) as described in the main text. Best fit values for the samples shown are 0.67 $\pm$ 0.01 and 0.72 $\pm$ 0.01 for $x$=0.24 and 0.25, respectively.}\label{fig_scaling}
\vspace{-0.4cm}
\end{figure*}

\subsection{Effective medium theory}

One possible scenario giving rise to critical phenomena and a possible temperature independent crossing point in magnetoresistance curves is percolation in a superconductor--non-superconductor composite. From a Symmetrical Effective Medium Theory \cite{EMT, stroud1}, the resultant resistivity $\rho_m$ of a composite where one of the components has $\rho_1=0$ (superconductor) and fills a fraction $p$ of the space, and the other has resistivity $\rho_2$ and fills a fraction $1-p$ is $\rho_m=(1/p_c)(p_c-p)\rho_2$ for $p<p_c$ and $\rho_m=0$ for $p>p_c$, where $p_c$ is the percolation threshold (which depends on the geometry of the space). In the case of BaPb$_{1-x}$Bi$_x$O$_3$, $\rho_2$ would represent the effective resistivity of the material in the normal state, given that the superconducting part associated with $\rho_1$ can become normal above a certain field or temperature. Therefore, if we assume that both $\rho_2$ and the superconducting fraction $p$ depend on temperature and field (to account for the effect mentioned before, as well as effects such as variations in $T_c$ and/or Josephson coupling between different superconducting regions), we can write $\rho_m(T,H)=(1/p_c)\left[p_c-p(T, H)\right]\rho_2(T,H)$. With this, the condition to have a temperature-independent crossing point at the critical field $H_c$ is $\partial\rho_m(T,H)/\partial T|_{H=H_c}=0$, or

\begin{equation}
\left. \frac{\partial p(T,H)}{\partial T}\right|_{H_c}\rho_2(T,H_c)-\left. \left[p_c-p(T,H_c)\right]\frac{\partial\rho_2(T,H)}{\partial T}\right|_{H_c}=0\label{percol}
\end{equation}

For this condition to be true over a range of temperatures requires fine tuning of the different parameters involved. From Fig. \ref{fig_rhoT}, we see that for samples near optimal doping, where a temperature-independent crossing point is observed, we have that both: the temperature dependence of $\rho$ is very weak, so we can possibly fulfill the condition $\partial\rho_2(T,H)/\partial T|_{H=H_c}\approx 0$, and, if the superconducting transition is driven by percolation, $p_c-p(T,H_c)<<1$. Therefore, the second term of Eqn. \ref{percol} is expected to be small for optimally doped samples. Hence, to get a temperature-independent crossing point requires that the first term in Eqn. \ref{percol} be very small, and that the two terms are finely tuned to give zero difference over an appreciable range of temperatures. The first requirement is possibly unphysical since it is hard to conceive a mechanism that would give rise to $\partial p(T,H)/\partial T|_{H=H_c}=0$, and the second requirement is at best rather unsatisfying. In particular, why should this fine tuning occur precisely at the composition that also happens to yield the maximum $T_c$?.


In addition, as we show below, the resistivity for fields on both sides of $H_c$ scales in a very specific way. Hence, not only is fine-tuning required to achieve a temperature-independent crossing point within a classical model, but this same level of fine-tuning is also required for all fields and temperatures close to $H_c$. As such, a purely classical model based on an effective medium theory appears to be inadequate to account for the observed phenomenology.

\subsection{Quantum phase transition}\label{QPT}

The temperature-independent crossing point found for compositions close to optimal doping implies the possibility of critical scaling. Indeed, quite remarkably, over the whole range of temperatures where $H_{crossing}$ shows convergence for these compositions, the magnetoresistance curves exhibit scaling of the form

\vspace{-0.3cm}
\begin{equation}
\rho(T,H)=\rho_c F\left(\frac{|H-H_{c}|}{T^{1/z\nu}}\right)\label{scal}
\end{equation}

where $\rho_c$ is temperature and field independent and varies from sample to sample and $F(y)$ is a universal function. Values of the critical exponents product $z\nu$ were obtained for each sample by two methods. Initial estimates were obtained from a linear fit to log$_{10}(d\rho/dH)|_{Hc}$ vs log$_{10}T$ (insets to Figs. \ref{fig_scaling}(a) and \ref{fig_scaling}(b)). Further refinement to these values was obtained by an iterative maximization (in H$_c$ and $z\nu$) of the correlation coefficient ($R^2$) calculated from a fitting of each part of the scaling function (for $H>H_c$ and $H<H_c$) to a generic function (ninth order polynomial). Initially $H_c$ was varied for values of $z\nu$ obtained by a best visual collapse of the data. With the optimal value of $H_c$, $z\nu$ was then varied to obtain the best $R^2$. The results of this procedure are summarized in the insets of Fig. \ref{fig_scaling}(a) and \ref{fig_scaling}(b) for two specific samples. For the upper branch ($H>H_c$), the correlation coefficient does not show a local maximum, whereas for the lower branch ($H<H_c$), it clearly shows one. The $z\nu$ value obtained by this method is the one that maximizes $R^2$ for the lower branch, and was confirmed to give the best visual collapse of the data. For the specific data sets shown in Fig. \ref{fig_scaling}, scaling of the full curves gave exponents $z\nu$ = 0.67 $\pm$ 0.01 and 0.72 $\pm$ 0.01 for $x$ = 0.24 and 0.25 respectively. Measurements were made for several crystals with the same nominal compositions. For all the samples analyzed, $z\nu$ takes values between 0.65 and 0.72, and the average value obtained is 0.69 $\pm$ 0.03.

Magnetic-field-tuned superconductor-insulator transitions (SITs) with temperature-independent crossing points have been widely studied in disordered 2D thin films of several different materials. A variety of critical exponent products have been reported for such magnetic-field-tuned transitions: $z\nu=1.3$ for InO$_x$ \cite{Steiner2, Steiner1} and MoGe,\cite{Yazdani1} $z\nu=1.5$ for La$_{2-x}$Sr$_x$CuO$_4$,\cite{Bollinger1} $z\nu=2.3$ for highly disordered InO$_x$,\cite{Steiner1} and $z\nu=0.7$ for $a$-Bi \cite{Markovic1, Parendo1} and $a$-Nb$_x$Si$_{1-x}$,\cite{Aubin1} these last two cases being similar to our observation for BaPb$_{1-x}$Bi$_x$O$_3$. In general, for a d-dimensional system undergoing a quantum phase transition, the expression for resistivity inside the quantum-critical region obtained by finite-size scaling has the form $\rho(T,y)\propto\xi^{d-2} F_{\rho}(L_{\tau}/\xi)$, where $\xi$ is the diverging order parameter correlation length, and $L_{\tau}$ is a real length associated with the correlation length $\xi_{\tau}\propto\xi^{z}$ in the imaginary time axis, introduced in the quantum-classical mapping of the transition by the relation $it/\hbar=1/k_B T$. Consequently, $L_{\tau}\propto(\hbar/k_B T)^{1/z}$, where $z$ is the dynamical critical exponent, introduces the temperature dependence of the scaling function, and $\xi\propto|y-y_c|^{-\nu}$, where $\nu$ is the correlation length critical exponent, introduces the dependence with the tuning parameter $y$. With this, $\rho(T,y)\propto T^{-(d-2)/z} F(|y-y_{c}|T^{-1/z\nu})$.   \cite{FFH2} This expression has the same form as Eqn. \ref{scal} for $d=2$, and a crossing point of different isotherms at the critical value $y_c$ is expected only for 2D systems. Such behavior has been extensively corroborated for the systems mentioned before.


Of course the principle objection to such a scaling in BaPb$_{1-x}$Bi$_x$O$_3$ is that our measurements are of bulk, 3-dimensional single crystals. The data could not be fitted to a 3D quantum phase transition (ie. with a temperature-dependent prefactor proportional to $T^{-1/z}$) with physically reasonable values of $z$ for any of the compositions studied. The apparent 2D scaling found for optimally doped samples might be a consequence of a ``hidden 2-dimensionality'' due to the nanostructure, perhaps in the form of a percolating network of superconducting sheets. Within such a scenario, subtle variation in the nanostructure of different samples might account for the low-temperature deviation from 2D scaling seen for some samples with $x$ = 0.24 and 0.25 (open symbols in Fig. \ref{fig_Hcrossing}), which presumably reflects the presence of a finite 3D coupling below some energy scale.  The absence of a well-defined crossing point for samples with $x \geq$ 0.28 would imply a deviation from 2D behavior, and could be related to the progressive reduction in the volume fraction of the tetragonal polymorph with increasing Bi concentration.\cite{Cava1} However, to obtain a temperature-independent crossing point in such a scenario would require all of the intergrowths to be aligned in a parallel fashion, so that the projection of the field onto each ``sheet'' is identical, which seems unlikely based on the pattern seen in Fig. \ref{fig_TEM}(b). Nevertheless, the streaks seen in the FFT in Fig. \ref{fig_HRTEM}(b) are suggestive of a preferred orientation of the polymorph intergrowths, but further analysis is required in order to determine if such scenario is feasible. Such a description is therefore equally unsatisfactory as the analysis based on the symmetrical effective medium theory, although the empirical observation of a temperature-independent crossing point and the associated scaling remain highly suggestive of a manifestation of some kind of quantum phase transition that mimics the behavior of a 2d system over a wide range of temperatures. 

It remains to be seen whether the phenomenology of the field-tuned superconductor-insulator transition in optimally doped BaPb$_{1-x}$Bi$_x$O$_3$ is truly driven by quantum critical behavior. However, such a scenario could naturally account for several specific features of this material, and the closely related case of Ba$_{1-x}$K$_x$BiO$_3$. Specifically, critical scaling following the form of Eq. (\ref{scal}) implies the presence of quantum fluctuations of the superconducting order parameter over a wide range of temperatures. Although quantum fluctuations have a negligibly small effect for clean superconductors with a moderate carrier density, reduction of the phase stiffness due to reduced superfluid density and poor screening in bad metals can lead to significant suppression of $T_c$ relative to the mean-field value.\cite{Kivelson1, Kivelson2} Such a scenario is consistent with the large and ``insulating''-like resistivity of BaPb$_{1-x}$Bi$_x$O$_3$. It is also consistent with the observation that the optimal $T_c$ occurs
at the composition at which strong localization effects become significant and the resistivity begins to develop an insulating temperature dependence.  Finally, this scenario could also provide a natural explanation for the long-standing question as to why BaPb$_{1-x}$Bi$_x$O$_3$ has a lower maximum $T_c$ ($\approx$ 12 K) than Ba$_{1-x}$K$_x$BiO$_3$ ($\approx$ 30 K), since, first, substitution on the Bi sites (which are responsible for the bands at the Fermi energy \cite{Franchini1, Yin1}) will effectively lead to a more disordered state than substitution on the Ba site, and second, the polymorphism in BaPb$_{1-x}$Bi$_x$O$_3$ also introduces disorder and localization effects, enhancing the effect of quantum fluctuations in the phase of the order parameter.\cite{John1} 

\section{Conclusions}

In summary, we have shown that optimally doped BaPb$_{1-x}$Bi$_x$O$_3$ has a remarkable magnetoresistance, exhibiting an apparently temperature-independent crossing point over a wide range of temperatures, and scaling reminiscent of a 2D superconductor-insulator field-tuned quantum phase transition. This surprising phenomenology cannot be accounted for satisfactorily by either a classical effective medium theory, or, given the 3D nature of the material, a simple field-tuned SIT in 2 dimensions. Given the complex nanostructure due to the dimorphic nature of BaPb$_{1-x}$Bi$_x$O$_3$ in this compositional range, our observations are suggestive that a full theoretical treatment must include aspects of both percolation and quantum-critical scaling.



\section{Acknowledgments}

The authors thank R. Jones for help with microprobe measurements, and S. A. Kivelson, A. Kapitulnik and N. P. Breznay for helpful discussions. This work is supported by AFOSR Grant No. FA9550-09-1-0583. The electron microscopy was performed at Ames Laboratory (M.J.K. and Y.Z.) and supported by the U.S. Department of Energy, Office of Basic Energy Science, Division of Materials Sciences and Engineering, under Contract No. DE-AC02-07CH11358.

\bibliography{1_BPBO_biblio}\vspace{-0.8cm}

\end{document}